\documentclass[10pt,twocolumn,letterpaper]{article}

\pdfoutput=1
\usepackage{iccv}
\usepackage{times}
\usepackage{epsfig}
\usepackage{graphicx}
\usepackage{amsmath}
\usepackage{amssymb}
\usepackage{authblk}
\usepackage[dvipsnames,table]{xcolor}

\usepackage[breaklinks=true,bookmarks=false]{hyperref}

\iccvfinalcopy 

\def\iccvPaperID{****} 

\begin{document}

\title{Cross-domain Food Image-to-Recipe Retrieval by Weighted Adversarial Learning}

\author[1]{Bin Zhu}
\author[2]{Chong-Wah Ngo}
\author[3]{Jingjing Chen}
\author[1]{Wing-Kwong Chan}

\affil[1] {City University of Hong Kong}
\affil[2] {Singapore Management University}
\affil[3] {Fudan University}




\maketitle

\begin{abstract}
      Food image-to-recipe aims to learn an embedded space linking the rich semantics in recipes with the visual content in food image for cross-modal retrieval. The existing research works carry out the learning of such space by assuming that all the image-recipe training example pairs belong to the same cuisine. As a result, despite the excellent performance reported in the literature, such space is not transferable for retrieving recipes of different cuisine. In this paper, we aim to address this issue by cross-domain food image-to-recipe retrieval, such that by leveraging abundant image-recipe pairs in source domain (one cuisine), the embedding space is generalizable to a target domain (the other cuisine) that does not have images to pair with recipes for training.
   With the intuition that the importance
   of different source samples should vary, this paper proposes two novel mechanisms for cross-domain food image-to-recipe retrieval,
   i.e., source data selector and weighted cross-modal adversarial learning. The former aims to select source samples similar to the target data and filter out distinctive ones for training. The
   latter is capable to assign higher weights to the source samples more similar to the target data and lower weights to suppress the distinctive ones for both cross-modal and adversarial learning.
   The weights are computed from the recipe features extracted from a pre-trained source model.
   Experiments on three different cuisines (Chuan, Yue and Washoku) demonstrate that the proposed method manages to achieve state-of-the-art performances in all the transfers.
\end{abstract}

\section{Introduction}
\label{sec:intro}
It is well-known that the progress in deep learning~\cite{dl} heavily relies on the availability of abundant labeled data. Examples of success include image classification~\cite{resnet,hu2018squeeze},
object detection~\cite{redmon2016you,he2017mask} and semantic segmentation~\cite{long2015fully}. However, provision of annotation for large-scale datasets like ImageNet~\cite{deng2009imagenet} is time-consuming
and cost-expensive. To alleviate this issue, domain adaptation~\cite{pan2009survey} aims to transfer the knowledge from a model trained with off-the-shelf abundant labeled data (source domain) 
for learning a new model which is able to generalize well on
a set of relevant data (target domain). As the new model is to deal with different data (target domain), transferring the model from source to target domain can easily lead to performance degradation.

This paper studies the problem of cross-domain food image-to-recipe retrieval. Specifically, the target domain contains only recipes during the training phase. The motivation of this setting is two-fold. On the one hand, it is not always easy to collect sufficient paired data for cross-modal learning, such as in countries with less developed social media (e.g., recipe-sharing websites), minority cuisines in certain areas and intellectual property problems in collecting data. On the other hand, most of the recipes on the sharing websites are not associated with food images. Take Recipe 1M~\cite{JNE} dataset as an example, only 33\% of recipes contain at least one image. The purpose of this paper is to leverage a source domain with abundant image-recipe pairs
to learn the embedding of the recipes such that the embedding is close to the food image of the recipes. Note that these images are not available during training. During testing time, given a query food
image, the task is to retrieve the recipe of the image. As discussed in~\cite{zhumm20}, by retrieving the recipe of a query image, ingredients can be directly extracted for nutrition estimation. Furthermore,
no labeling efforts are required in contrast to classification approaches~\cite{bossard2014food,chen2016deep,min2019ingredient, chen2020study} that annotate food and ingredients labels for each image. 

\begin{figure*}
   \begin{center}
   \includegraphics[scale=0.24]{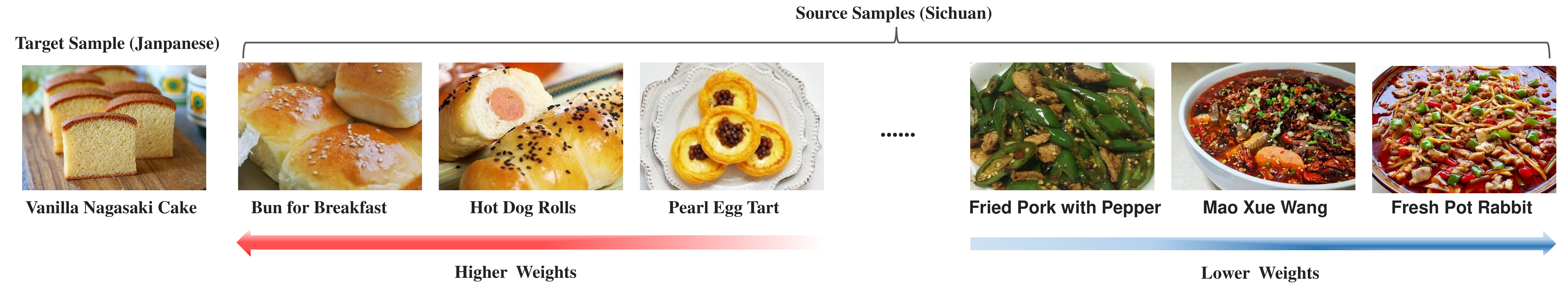}
   \end{center}
      \caption{Examples showing Sichuan cuisine as source domain and Japanese cuisine as target domain. The proposed method assigns higher
    weights to the source samples more similar to the target data and lower weights to suppress the distinctive ones. The title of a dish is shown below each food image.}
   \label{fig:model}
\end{figure*}

The existing cross-modal recipe retrieval works~\cite{JNE,jj18,carvalho2018cross, R2GAN,ACME,MCEN, zhu2021learning, salvador2021revamping, papadopoulos2022learning, xie2022cross} assume that
the data distributions of the training and testing sets are the same. In contrast, this paper deals with the problem that source and target data derives from different domains. 
In cross-domain food image-to-recipe retrieval, this refers to the source and target domains having different cuisines of dishes. The domain gap between different cuisines
may result from different ingredient composition, cutting and cooking methods, food culture and so on. For instance, Mexico cuisine tends to be spicy due to the frequent use of ``chilli''. While French cuisine
is less spicy, and the common ingredients in French dishes are ``cheese'' and ``truffle''. The problem is thus challenging due to domain gap (i.e., different cuisines), the use of multiple modalities between
source and target domains as well as no pairing information to be leveraged in the target domain.

In the literature, most of the existing works pay attention to single-modal domain adaptation~\cite{tzeng2014deep,long2015learning, WMMD, DANN, tzeng2017adversarial, WPDA, WAAN, wang2020progressive, cui2020gradually}.
The domain gap is bridged by either metric measurement~\cite{tzeng2014deep,long2015learning, WMMD} or adversarial learning~\cite{GAN, DANN, tzeng2017adversarial, WPDA, WAAN, wang2020progressive, cui2020gradually}.
Recently, a few works~\cite{qi2018unified, huang2018mhtn, zhumm20} explore domain transfer in cross-modal perspective, which enrich the modalities to text, audio, image and so on. The emphasis of these works is
different, ranging from predicting class label of target cross-modal pairs using labeled source pairs~\cite{qi2018unified}, labeling target data in multiple modalities using single modality labeled source data~\cite{huang2018mhtn}, to
pairing partial target data using complete paired source data in food domain~\cite{zhumm20}. As the recipe contains rich textual information, including title, ingredients list and cooking steps describing the cooking
procedure, the improvement achieved by recipe is more significant than image alone~\cite{zhumm20}. Therefore, this paper focuses on the case of the target domain containing only recipes. Furthermore, a domain classifier is used to align the source and target distributions in~\cite{zhumm20} without considering
the unique properties (e.g., ingredients and cooking styles) within cuisines, thus the distinctive source samples could introduce noise during training, which could limit the transfer performance.

This paper addresses the limitation by proposing data selection and re-weighting schemes. Specifically, we claim that different weights should be assigned to the source data to align the data distribution with target domain,
based on the similarities with the target data. The intuition lies in that different samples demonstrate various degrees of properties peculiar to a certain domain (cuisine). Hence, the unique source samples should be
suppressed with lower weights and the more similar ones deserve more importance for feature alignment. To achieve this, source batch selector (SBS) is proposed to pick the samples more similar to the target
data to form the source batch for training. Additionally, weighted adversarial cross-modal learning is proposed to re-weight the cross-modal and adversarial learning with weighted triplet loss and weighted adversarial loss respectively,
where the weights are computed from the source and target features extracted from a pre-trained source model.

\section{Related Work}
Domain adaptation addresses the problem of knowledge transfer from a source domain to a target domain~\cite{pan2009survey}. Due to the domain gap in data distribution between the source and target domains, the transfer may adversely degrade performance.
In the literature, most of the works attempt to reduce the impact of domain shift by bridging the gap between source and target data in feature space.
On the one hand, domain discrepancy is directly computed by metrics like maximum mean discrepancy (MMD)~\cite{tzeng2014deep,long2015learning,WMMD}. An adaptation layer with MMD in~\cite{tzeng2014deep} is used to
learn domain-invariant deep features. \cite{long2015learning} extends this work by proposing the multi-kernel variant of MMD which selects optimal kernels to improve the transferability of features. In~\cite{WMMD},
weighted MMD introduces class-specific weights for source domain to address the issue of class bias across domains.

On the other hand, some other works focus on learning domain-invariant features adversarially by using a domain classifier~\cite{DANN, tzeng2017adversarial, WPDA, WAAN, wang2020progressive, cui2020gradually}.
Domain adversarial neural network (DANN)~\cite{DANN} employs a domain classifier to distinguish whether the input is from the source or target domains. The feature extractor is connected to the domain classifier by a
gradient reversal layer which reserves the gradients during back-propagation. In~\cite{tzeng2017adversarial}, the training is divided into two stages, in the first stage, a source model is firstly learnt by only using
the source data, then a discriminator along with the pre-trained source model is used for adversarial adaptation to learn the target model. Weighted-based adversarial methods are also explored in~\cite{WPDA, WAAN}.
Re-weighted adversarial adaptation network (RAAN)~\cite{WAAN} proposes to re-weight source domain label distribution to match the target label distribution by minimizing Earth Mover distance to adapt classifier with adversarial learning.
In~\cite{WPDA}, an extra domain classifier is used to calculate the weights for source samples. Then, the weights together with source and target data are fed into the second domain classifier for adversarial training. In~\cite{cui2020gradually}, gradually vanishing bridge mechanism is designed to produce an intermediate domain between source and target domains which enhances the balance of adversarial training.
Progressive adversarial network (PAN)~\cite{wang2020progressive} is presented which integrates curriculum learning~\cite{bengio2009curriculum} and adversarial learning for fine-grained domain adaptation.
In general, this paper belongs to weighted-based adversarial methods. Different from~\cite{WPDA, WAAN}, we focus on cross-domain food image-to-recipe retrieval rather than visual domain adaptation. Furthermore, the weights
are produced from pre-trained source model in this paper instead of an extra discriminator~\cite{WPDA} or minimizing Earth Mover distance~\cite{WAAN}. We also propose source batch selector to filter source samples and weighted scheme
particularly for cross-modal adversarial learning.

\begin{figure*}
   \begin{center}
   \includegraphics[scale=0.31]{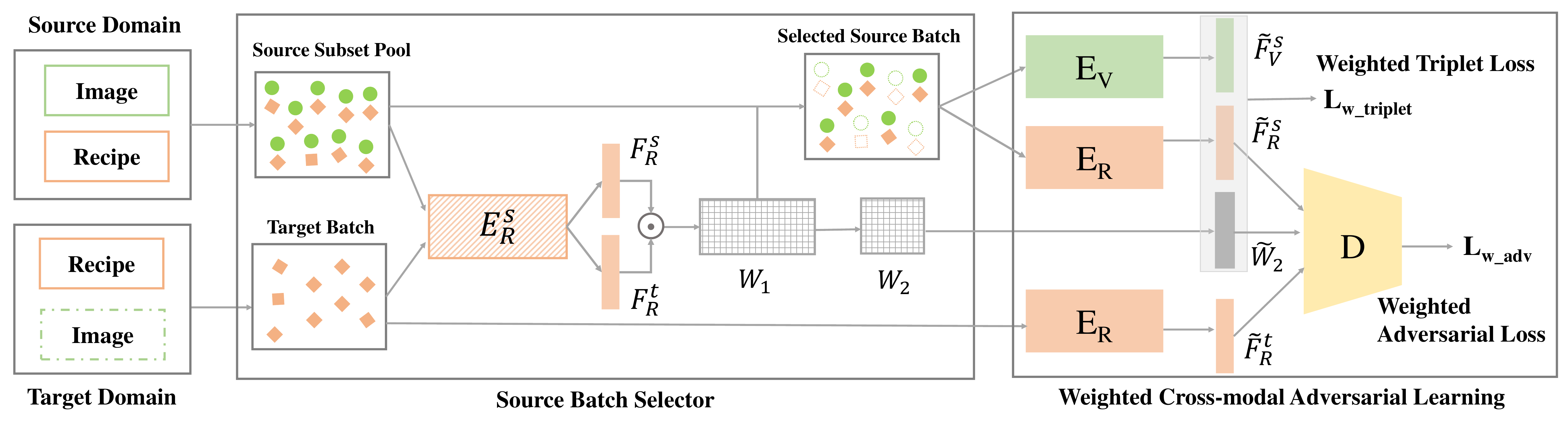}
   \end{center}
      \caption{The overview of the proposed method. The source batch selector aims to filter out distinctive source samples from the target data, and the weighted cross-modal adversarial learning assigns higher
      weights to the more similar samples for source and target distribution alignment in both cross-modal and adversarial learning. The matrixes $W_1$ and $W_2$ are both produced by a pre-trained source model ($E^s_R$)
      which is frozen during training procedure.}
   \label{fig:model}
\end{figure*}

Most of the existing domain adaptation methods only consider single modality. Cross-modal transfer is rarely studied~\cite{qi2018unified, huang2018mhtn, zhumm20}. \cite{huang2018mhtn} focuses on transferring knowledge
from a single modality source domain to a cross-modal target domain. For example, the source domain is a set of labeled images while target could be unlabeled text, video or audio. A modal-adversarial hybrid transfer network is proposed
to semantically align the features across modalities by equipping MMD for task-specific layers and domain classifier for adversarial learning simultaneously. In~\cite{qi2018unified}, both the source and target data contains image-audio pairs and only
source samples have class labels. The goal is to label the target pairs using the knowledge learnt from source domain. The features extracted from different modalities are used to compute attention scores and then fused together
for classification. Three domain classifiers are employed to discriminate modality, attention scores and fused features respectively for feature alignment.

The most relevant work to this paper is cross-domain cross-modal food transfer (CCFT)~\cite{zhumm20}, which studies cross-modal transfer in food domain.
CCFT considers transferring knowledge across different modalities and from complete to partial data. For example, while the source domain contains both images and recipes for training, the
target domain contains only recipes or even simply the titles of dishes for training. While the problem scenario is interesting, CCFT merely exploits to use domain classifiers for recipe and image features
to learn domain-invariant features for cross-modal and incomplete data transfer. Due to different cuisines showing certain unique properties, there may exist part of samples in the source domain that do not contribute to the transfer but increase
noise for training. By equally aligning all the features from source and target domain, the transfer performance of CCFT is limited. This paper addresses the limitation by proposing data selection and re-weighting mechanisms to filter out distinctive source samples
for training and assign different weights to each source samples based on similarity to target domain for cross-modal and adversarial learning. The recent work~\cite{zhu2022cross} explores mixup by exchanging the recipe sections between source and target domains for cross-lingual recipe adaptation. Since the language gap is still extremely challenging to conquer without machine translation~\cite{zhu2022cross}, we focus on recipe adaptation in the same language. 

\section{Method}
By leveraging a source domain with abundant training data, the goal is to perform image-to-recipe retrieval on a target domain. In this paper, we focus on the scenario that
only recipe exists in the target domain. It has been reported in~\cite{zhumm20} that,
due to the provision of textually rich description, recipe is more helpful than image in boosting the domain transfer performance. Food image alone, without pairing with a recipe, offers limited performance.
Denote source domain as $\mathcal{D}^s=\{(r^s_i, v^s_i)\}^{N^s}_{i=1}$ with $N^s$ recipe-image pairs, where $r^s_i$ is a source recipe and $v^s_i$ is the paired dish image. Meanwhile, target domain contains a set of recipes, denoted as $\mathcal{D}^t=\{(r^t_j)\}^{N^t}_{j=1}$,
where $r^t_j$ is a target recipe. The superscripts $s$ and $t$ refer to the source and target domains respectively. It is worth noting that the recipes in both source and target domains are assumed to be complete, in other words, each recipe contains three sections (i.e., title, ingredients list and cooking steps).

Figure~\ref{fig:model} depicts the architecture network of the proposed weighted cross-modal adversarial learning model. With the intuition that the samples should not be treated equally in cross-domain food image-to-recipe retrieval, two key mechanisms are designed: one is source batch selector (SBS, Section~\ref{subsec:SBS}), which is able to cherry-pick source samples more similar to the target domain for training. The other is weighted cross-modal adversarial learning (Section~\ref{subsec:WAL}), which
enables the model to weighted align the source and target distribution not only in cross-modal learning but also in adversarial learning by assigning different weights to the selected source batch data.

\subsection{Source Batch Selector} \label{subsec:SBS}

The aim of source batch selector is to filter out ``bad'' source samples which are far away from target data from the very beginning. We claim that the model should pay more attention to the data
that is similar to the target data rather than the unique or distinctive samples in the source domain. As shown in Figure~\ref{fig:model}, a source subset pool with $n^s$ recipe-image pairs
and a target batch $n^t$ recipes are randomly sampled from source and target data respectively (let $n^s\geq n^t$). The recipe encoder $E^s_R$ pre-trained on source data is subsequently employed to obtain both source and target recipe features
denoted as $F^s_R=\{(f_{r^s_i})\}^{n^s}_{i=1}$ and $F^t_R=\{(f_{r^t_j})\}^{n^t}_{j=1}$ respectively, where $f_{r^s_i}$ and $f_{r^t_j}$ are the features of $r^s_i$ and $r^t_j$ respectively. Note that the weights of $E^s_R$ are frozen during the entire training procedure.
A matrix $W_1 \in \mathbb{R}^{n^t\times n^s}$ that captures the pairwise similarity between recipes in the target batch and recipes in the source subset pool is then computed as follows:

\begin{equation}
   \begin{aligned}
       W_1(j,i) = \frac{f_{r^t_j} \cdot f_{r^s_i}}{{\big\| f_{r^t_j} \big\|} {\big\|f_{r^s_i} \big\|}}.
   \end{aligned}
   \label{eq:cos}
\end{equation}

Based on $W_1$, a new source subset is formed by collecting the top $K$ most similar samples in the source pool for each target recipe, where $K$ is a hyperparameter. In practice, we set $K=2$ when $n^s\geq 2n^t$. Hence, the remaining
samples in the new subset are somewhat similar to at least one target sample and the distinctive samples are excluded from training. With this, the source batch can be randomly selected from the new subset to match the size of the target batch for
further weighted cross-modal adversarial learning. Along with this, a matrix $W_2 \in \mathbb{R}^{n\times n}$ is derived from $W_1$, which represents the similarity between recipes of the selected source batch
and the target batch, where $n$ is the batch size.

\begin{table*}
   \caption{Cross-domain food image-to-recipe retrieval performance comparison in terms of MedR. Note that ``w/o w\_triplet'' and ``w/o w\_adv'' refer to using the original triplet and adversarial losses without re-weighting respectively. The oracle model shows the upper-bound performance by fully supervised training with both source and target paired data.
   The last column is the average of all the transfers. ``C'', ``Y'' and ``W'' represent ``Chuan'', ``Yue'' and ``Washoku'' respectively.}
   \label{tab:performance}
   \begin{center}
   \begin{tabular}{l|c|c|cc|cc|c}
   \hline
   Methods & C$\rightarrow$W (1K) & Y$\rightarrow$W (1K) & C$\rightarrow$Y (1K) & C$\rightarrow$Y (5K) & Y$\rightarrow$C (1K) & Y$\rightarrow$C (5K) & Avg.\\
   \hline
   Random & 500 & 500 & 500 & 2500 & 500 & 2500 & 1,166.7\\
   Source-only~\cite{zhumm20} & 33.6 & 19.2 & 4.8 & 19.6 & 12.5 & 54.7 & 24.1\\
   IWAN~\cite{WPDA} & 33.0 & 18.8 & 4.7 & 18.5 & 12.2 & 54.4 & 23.6\\
   CCFT~\cite{zhumm20} & 11.7 & 15.7 & 4.2 & 16.6 & 8.8 & 38.6 & 15.9\\
   \hline
   \hline
   Proposed w/o SBS & 11.5 & 14.3 & 4.2 & 16.5 & 8.6 & 37.5 & 15.4\\
   Proposed w/o w\_triplet & 11.3 & 13.7 & 4.1 & 16.4 & 8.6 & 36.1 & 15.0\\
   Proposed w/o w\_adv & 11.2 & 13.3 & 4.1 & 16.3 & 8.3 & 35.9 & 14.9\\
   \textbf{Proposed} & \textbf{10.6} & \textbf{13.1} & \textbf{4.0} & \textbf{16.1} & \textbf{7.9} & \textbf{35.2} & \textbf{14.5}\\
   \hline
   \hline
   \rowcolor{black!40}
   Oracle (upper-bound) &  4.9 & 8.0 & 2.7 & 8.7 & 2.0 & 6.7 & 5.5\\
   \hline
   \end{tabular}
   \end{center}
 \end{table*}

\subsection{Weighted Cross-modal Adversarial Learning} \label{subsec:WAL}
The basic idea of weighted cross-modal adversarial learning is to assign higher weights to source samples more similar to target data and suppress the dissimilar ones with lower weights. As shown
in Figure~\ref{fig:model}, the source and target recipes are fed into a sharing weights learnable recipe encoder $E_R$ simultaneously.
The source and target recipe features extracted by $E_R$ are represented as $\tilde{F}^s_R=\{(\tilde{f}_{\tilde{r}^s_i})\}^{n}_{i=1}$ and $\tilde{F}^t_R=\{(\tilde{f}_{\tilde{r}^t_j})\}^{n}_{j=1}$ respectively,
where $\tilde{f}_{\tilde{r}^s_i}$ is the feature of $\tilde{r}^s_i$, $\tilde{f}_{\tilde{r}^t_j}$ is the feature of $\tilde{r}^t_j$, and $n$ is the batch size.
In addition, the source image features are extracted by a learnable image encoder $E_V$, denoted as  $\tilde{F}^s_V=\{(\tilde{f}_{\tilde{v}^s_i})\}^{n}_{i=1}$, where $\tilde{f}_{\tilde{v}^s_i}$ is the feature of $\tilde{v}^s_i$.

To obtain a one-dimensional weights vector $\tilde{W_2} \in \mathbb{R}^n$ used for model training, the similarity matrix $W_2$ is further processed as follows:

\begin{equation}
   \begin{aligned}
       &\tilde{W}_2(i) \leftarrow \sum_{j=1}^{n} W_2(i,j), \\
       &\tilde{W}_2(i) \leftarrow \frac{\tilde{W}_2(i)-\min(\tilde{W}_2)}{\max{(\tilde{W}_2)-\min(\tilde{W}_2)}},\\
       &\tilde{W}_2(i) \leftarrow \frac{n \tilde{W}_2(i)}{\sum_{i=1}^n\tilde{W}_2(i)}.
   \end{aligned}
   \label{eq:w2}
\end{equation}

After summing up across the target batch, normalizing and scaling, the weights vector $\tilde{W}_2$ represents the importance of each source sample with regard to the target data in the batch.

The proposed model is trained by the following objective function:

\begin{equation}
   \begin{aligned}
       \mathcal{L} = \mathcal{L}^s_{w\_triplet} + \beta \mathcal{L}_{w\_adv} + \lambda \mathcal{L}^s_{reg},
   \end{aligned}
   \label{eq:loss}
\end{equation}
where $\mathcal{L}^s_{w\_triplet}$ is the weighted triplet loss, $\mathcal{L}_{w\_adv}$ is the weighted adversarial loss and $\mathcal{L}_{reg}$ is the regularization loss. $\beta$ and $\lambda$ are trade-off hyperparameters.
The superscript $s$ indicates the losses can be only applied in the source domain.

Similar to the recent works~\cite{MCEN, ACME, R2GAN} in cross-modal recipe retrieval, triplet loss is adopted to push positive recipe-image pairs approaching and negative ones apart. The difference is that we
re-weight the source data when computing the triplet loss for each sample to reduce the gap between source and target domains. The weighted triplet loss is formalized as follows:
\begin{equation}
   \begin{aligned}
      \mathcal{L}^s_{w\_triplet} = &\tilde{W_2}[d(\tilde{f}_{\tilde{r}^s_a},\tilde{f}_{\tilde{v}^s_p}) - d(\tilde{f}_{\tilde{r}^s_a},\tilde{f}_{\tilde{v}^s_n}) + \alpha]_+ \\
      &+ \tilde{W_2}[d(\tilde{f}_{\tilde{v}^s_a},\tilde{f}_{\tilde{r}^s_p}) - d(\tilde{f}_{\tilde{v}^s_a},\tilde{f}_{\tilde{r}^s_n}) + \alpha]_+,
   \end{aligned}
   \label{eq:triplet}
\end{equation}
where $d(\cdot,\cdot)$ represents cosine similarity distance, subscripts $a$, $p$ and $n$ refer to anchor, positive and negative samples respectively, and the hyperparameter $\alpha$ is the margin.
$\mathcal{L}^s_{w\_triplet}$ contains two terms due to the anchor can be either recipe or image features.

To align the source and target distribution, a domain discriminator $D$ is employed. With the recipe features of source $\tilde{F}^s_R$ and target $\tilde{F}^t_R$ as input, the discriminator
is trained adversarially with the recipe encoder $E_R$ to predict the domain label. Specifically, $E_R$ is trained to produce recipe features which are indistinguishable from $D$ while $D$ maximizes
the probability of correct prediction. The weighted adversarial loss is defined as:
\begin{equation}
   \begin{aligned}
      \mathcal{L}_{w\_adv} = &\tilde{W_2}\mathbb{E}_{\tilde{F}^s_{R}\sim p_{R}^s}[\log{D(\tilde{F}_{R}^s)}]\\
      &+\tilde{W_2}\mathbb{E}_{\tilde{F}_{R}^t\sim p_{R}^t}[\log{(1-D(\tilde{F}_{R}^t)}].
   \end{aligned}
   \label{eq:adv}
\end{equation}

Following~\cite{zhumm20}, a regularizer $\mathcal{L}^s_{reg}$ is also used, which consists of two parts: multi-label classification loss and reconstruction loss. The former is produced by ingredients prediction
error from the image features and the latter comes from the difference between the real images and the reconstructed images by recipe features.

\section{Experiments}
\subsection{Experimental Settings}
\textbf{Datasets.}
The experiments are conducted on three datasets, which correspond to three cuisines, i.e., Sichuan (``Chuan''), Cantonese (``Yue'') and  Japanese (``Washoku'')~\cite{zhumm20}. Among them,
Chuan contains 42,797 recipes and 155,750 images, Yue contains 27,256 recipes and 119,758 images, and Washoku contains 9,626 recipes and 48,485 images. Note that all the recipes are in Chinese.
A list of 1,635 common ingredients across the three cuisines is also provided by~\cite{zhumm20}.

\textbf{Evaluation Metrics.} Following~\cite{zhumm20}, the performance is evaluated under the settings of cross-modal recipe retrieval~\cite{JNE}. Given a food image as query, the task is to retrieve
the relevant recipes.
Median rank (MedR) and recall rate at top K (R@K)
are employed to evaluate the transfer performance. MedR is the median position among the ranks of true positives for all the testing queries. R@K indicates the percentage of query images that the
corresponding recipes are ranked at Top K returned results. Hence, a lower MedR and higher R@K signify a better transfer model. Note that the experiments are repeated 10 times and the mean results
of MedR and R@K are reported.

\textbf{Baselines.} We compare the proposed method against two state-of-the-art approaches~\cite{WPDA,zhumm20} and two baselines, random and source-only. Baseline random is to randomly rank the recipes
given the query images. Baseline source-only~\cite{zhumm20} refers to training a standard cross-modal recipe retrieval model by only using the data from source domain and directly applying to the target
domain for testing. Furthermore, a weighted-based adversarial domain adaptation method~\cite{WPDA} and an adversarial-based cross-domain food image-to-recipe retrieval method~\cite{zhumm20} are also compared. Importance Weighted
Adversarial Net (IWAN)~\cite{WPDA} learns the domain-invariant features by dual discriminators, one is used to produce weights and the other is used for distinguishing between source and target domains.
The source model is pre-trained beforehand and the its parameters are not updated during the adversarial training between target model and the second discriminator. To the best of our knowledge,
CCFT~\cite{zhumm20} is the only existing work especially for cross-domain food image-to-recipe retrieval. In the case of only recipe existing in the target domain, CCFT adopted one domain classifier to adversarially
align the recipe features between source and target domains. The performance of oracle model is also presented as upper-bound for reference, which is trained by using all the paired data in source and target domains.

\textbf{Implementation Details.} The backbone of image encoder $E_V$ is based on ResNet-50~\cite{resnet} and initialzed with the weights pre-trained on ImageNet~\cite{deng2009imagenet}. The last layer of
ResNet-50 is replaced with a fully-connected layer with 1024-dimensional output. Same with the image features, the dimension of recipe features is also set to 1024. The word embedding is obtained via word2vec~\cite{word2vec} 
and the dimension is set to be 300. Similar with~\cite{JNE,zhumm20},
the recipe encoder $E_R$ firstly extracts title, ingredients and cooking steps raw features respectively. Bi-directional LSTM~\cite{lstm} is employed for title and ingredient encoding, while cooking steps
is processed by hierarchical LSTM considering the lengthy procedure text. Then, the three features are concatenated and fed into a fully connected layer for feature transformation to output the 
recipe features. Discriminator $D$ is a domain classifier with a three-layer perceptron which outputs the probability of source and target domains.

The model is trained by using Adam optimizer ~\cite{adam} with a batch size of 32 for all the experiments. The size of source subset pool is selected based on the transfer performance, which is detailed in Section~\ref{sec:ps}. 
The initial learning rate is set to be 0.0001 and we stop training until the model convergence. The trade-off hyperparameters $\beta$ and $\lambda$ in Equataion~\ref{eq:loss} are set to be 0.01 and 0.002 respectively.
Meanwhile, the margin in Equation~\ref{eq:triplet} is set to be $\alpha$=0.3. All the models are implemented using Pytorch~\cite{pytorch} framework.

\begin{figure*}
   \begin{center}
   \includegraphics[scale=0.50]{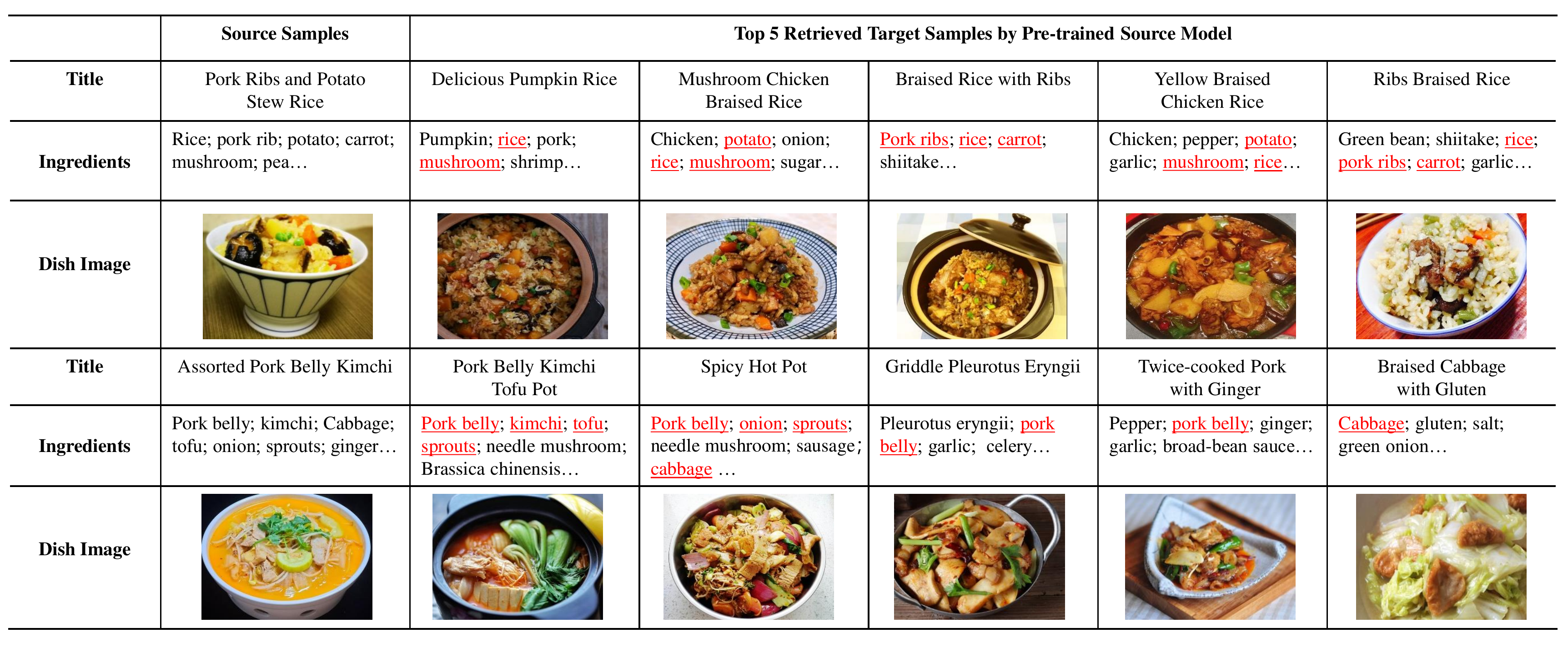}
   \end{center}
      \caption{Examples of top 5 retrieved target samples using source recipes as query via pre-trained source model for feature extraction in Y$\rightarrow$C transfer. Ingredients that appeared in the source recipe are underlined and highlighted in red.}
   \label{fig:effect_source}
\end{figure*}

\begin{figure*}
    \begin{center}
    \includegraphics[scale=0.23]{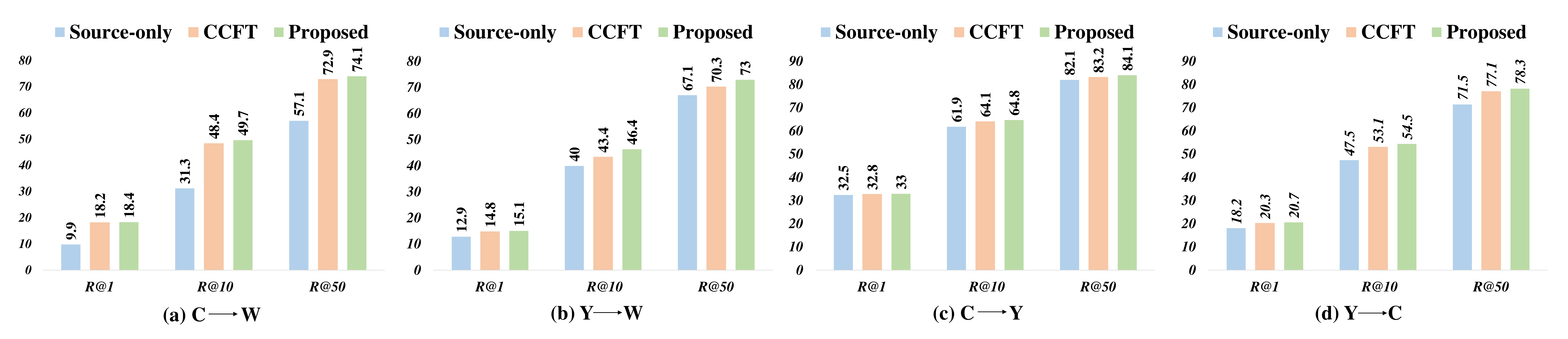}
    \end{center}
       \caption{Cross-domain food image-to-recipe retrieval performance comparison in terms of R@K in 1K test set.}
    \label{fig:recall}
 \end{figure*}

\subsection{Effectiveness of the Pre-trained Source Model}

As the pre-trained source model plays a key role in sample selection, we firstly analyze its effectiveness. Figure~\ref{fig:effect_source} shows examples of taking the source recipes (Yue) as queries to
retrieve the recipes in the target domain (Chuan), by using the recipe features extracted from the pre-trained source model. Note that the food images are shown for visualization purpose and are not involved
in retrieval. The ingredients of a target recipe are underlined and highlighted in red if appeared in the query recipe ($3_{rd}$ and $6_{th}$ rows).
As observed in the first query (rows 2-4), the top 5 retrieved recipes share some of the main ingredients as the query.
For instance, ``rice'' appears in all the recipes, ``mushroom'' in the recipes ``Delicious Pumpkin Rice'',
``Mushroom Chicken Braised Rice'' and ``Yellow Braised Chicken Rice'', as well as ``pork ribs'' in recipes ``Braised Rice with Ribs'' and ``Ribs Braised Rice''. Similar to the second query (rows 5-7),
the first four recipes contain the main ingredient ``pork belly'' while the last recipe ``Braised Cabbage with Gluten'' contains ``cabbage'' as the query. Furthermore, observed from the visual perspective,
the ingredients in different domains or even different dishes belonging to the same domain exhibit high variance in shape, texture and color due to the different cooking and cutting methods. Nevertheless, the high-level semantics are aligned and the pre-trained
source model is capable to mine the underlying information.

The MedR performances achieved by the pre-trained source model of Y$\rightarrow$C transfer are 12.5 and 54.7 in 1K and 5K test sets respectively, as shown in Table~\ref{tab:performance}. Despite not being perfect, the model indeed provides useful hints and evidences between
the correlation of source and target domains according to the analysis above. In addition, the model is consistent during the training procedure as its parameters are not updated. Consequently, we believe that the pre-trained source model
is decent and reasonable to measure the similarity of the source and target recipes.

\begin{figure*}
   \begin{center}
   \includegraphics[scale=0.338]{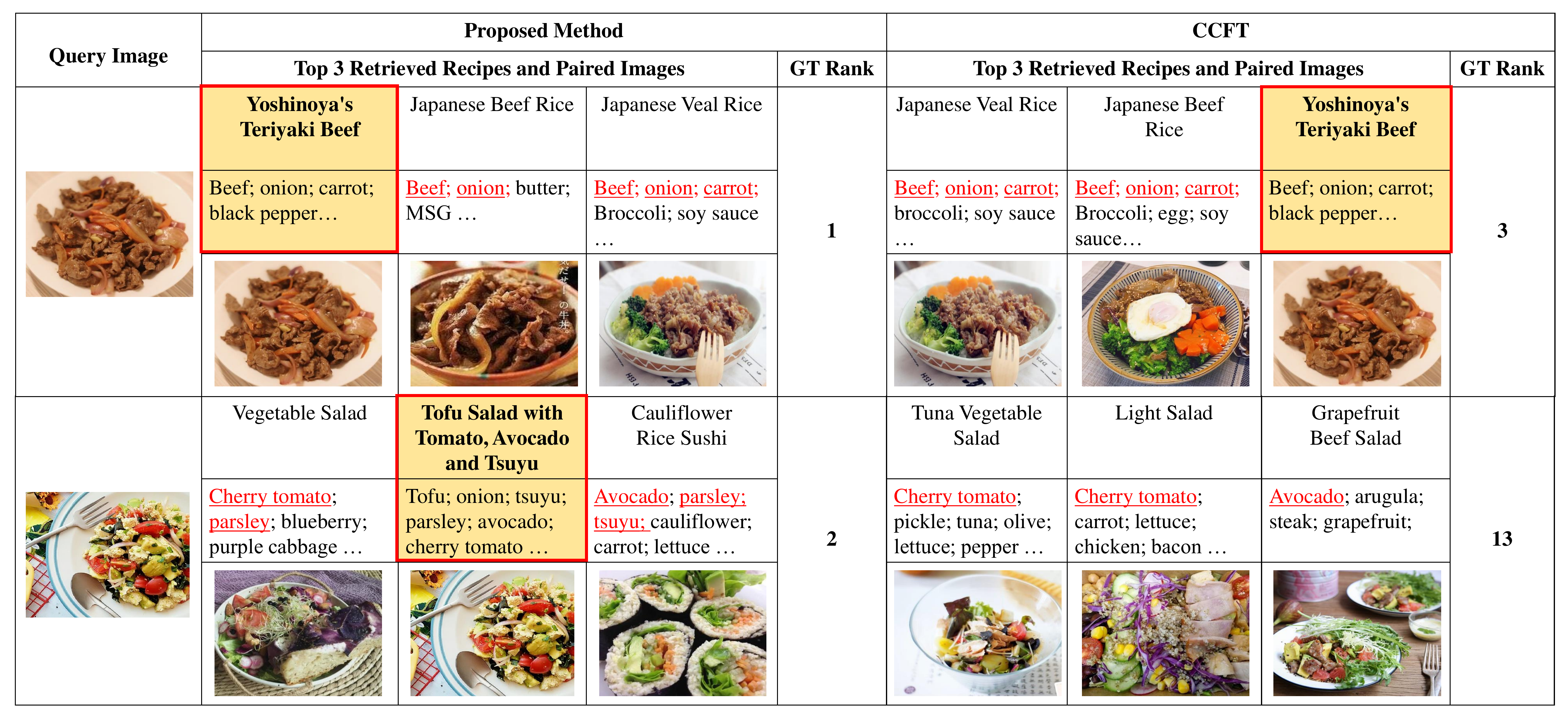}
   \end{center}
      \caption{Examples showing the image-to-recipe retrieval performance comparison between CCFT~\cite{zhumm20} and the proposed method in the target domain via Y$\rightarrow$W transfer. The ground truth (GT) recipe of a query image is highlighted
      with red bounding box and the ingredients that appeared in a GT recipe are underlined and highlighted in red.}
   \label{fig:comparison}
\end{figure*}

\subsection{Performance Comparison}

\textbf{Quantitative Results.}
Table~\ref{tab:performance} shows the performance comparison of the proposed method against the baselines in terms of MedR. First of all, the proposed method manages to outperform all the baseline methods among all the transfers.
Second, the improvements exhibit large variance for different transfers. When comparing to the pioneering work CCFT~\cite{zhumm20} for cross-domain food image-to-recipe retrieval, MedR is boosted with a large margin in C$\rightarrow$W,
Y$\rightarrow$W and Y$\rightarrow$C, while the improvements for C$\rightarrow$Y are relatively tiny, which only achieved by 0.2 and 0.5 ranks in 1K and 5K test sizes. Nevertheless, the MedR is improved by 8.8\% on average,
from 15.9 to 14.5. Third, the proposed method significantly outperforms one of the state-of-the-arts weighted-based adversarial method IWAN~\cite{WPDA}. Comparing to Source-only model, IWAN only achieves
slight improvements, which shows limited applicability for cross-domain food image-to-recipe retrieval.

R@K performance is also presented in Figure~\ref{fig:recall} with the comparison of Source-only, CCFT and the proposed model. Both the proposed model and CCFT manage to boost the performance with a large margin in comparison with Source-only model
by only using unpaired recipes. Meanwhile, the proposed model outperforms CCFT in all transfers for all R@K. It is worth noting that the improvements are more noticeable in R@10 and R@50 than R@1. In other words, the rank gains are mostly from
the middle and later instead of the foremost positions. For example, in Y$\rightarrow$W transfer (Figure~\ref{fig:recall} (b)), R@10 and R@50 are improved by 3\% and 2.7\% respectively but only 0.3\% for R@1. To sum up,
the ability to distinguish the fine-grained recipes in the top positions is still limited without involving the paired information for cross-domain food image-to-recipe retrieval.

Despite the improvements achieved by the proposed method, a noticeable gap can be observed compared to the oracle model. Since the oracle model is trained using paired data from source and target domains in full supervision, the performance is the upper-bound for all the models. Interestingly, the gaps between CCFT and the oracle models are quite diverse for different transfers, for example, the gap between oracle and CCFT in C$\rightarrow$W is 6.8 while that of C$\rightarrow$Y is 1.5. It is because the domain shift and the number of training data in different cuisines vary. Similarly, the margins between our proposed method and CCFT are also different. 
 
\textbf{Qualitative Results}. Figure~\ref{fig:comparison} shows two representative examples in Y$\rightarrow$W transfer that the proposed method improves the rank over CCFT. Taking Washoku food image as query, top 3
retrieved recipes by the two methods are shown, along with the paired food images. The ground truth (GT) recipe is highlighted with bounding box and the ingredients of the counterpart recipes are underlined and highlighted in red if appeared in GT.
Note that the target images are never seen during the model training. In the first example (rows 3-5), both methods are able
to rank the GT recipe within top 3, where the rank positions for the proposed method and CCFT are 1$_{st}$ and $3_{rd}$ respectively. Both of the top 3 retrieved recipes are very similar
to the GT recipes except for tiny differences in ingredient composition. Unlike the proposed method, CCFT fails to differentiate the recipes ``Japanese Veal Rice'' and ``Japanese Beef Rice'' with ingredients ``broccoli'' which does not exist in the GT recipe.
In the second example (rows 6-8), the rank of the proposed method is improved by 11 positions comparing to CCFT, from 13 to 2. The number of shared ingredients with GT recipe in the top 3 retrieved recipes of CCFT is obviously less than that of the proposed method.
Nevertheless, the top 1 retrieved recipe ``Vegetable Salad'' which contains ``cherry tomato'' and ``parsley'' is still ranked ahead one position than GT recipe by the proposed method, which shows the challenge of the task in differentiating the recipes with subtle difference in ingredient composition.

\subsection{Parameter Selection and Ablation Study} \label{sec:ps}
\begin{figure}
   \begin{center}
   \includegraphics[scale=0.35]{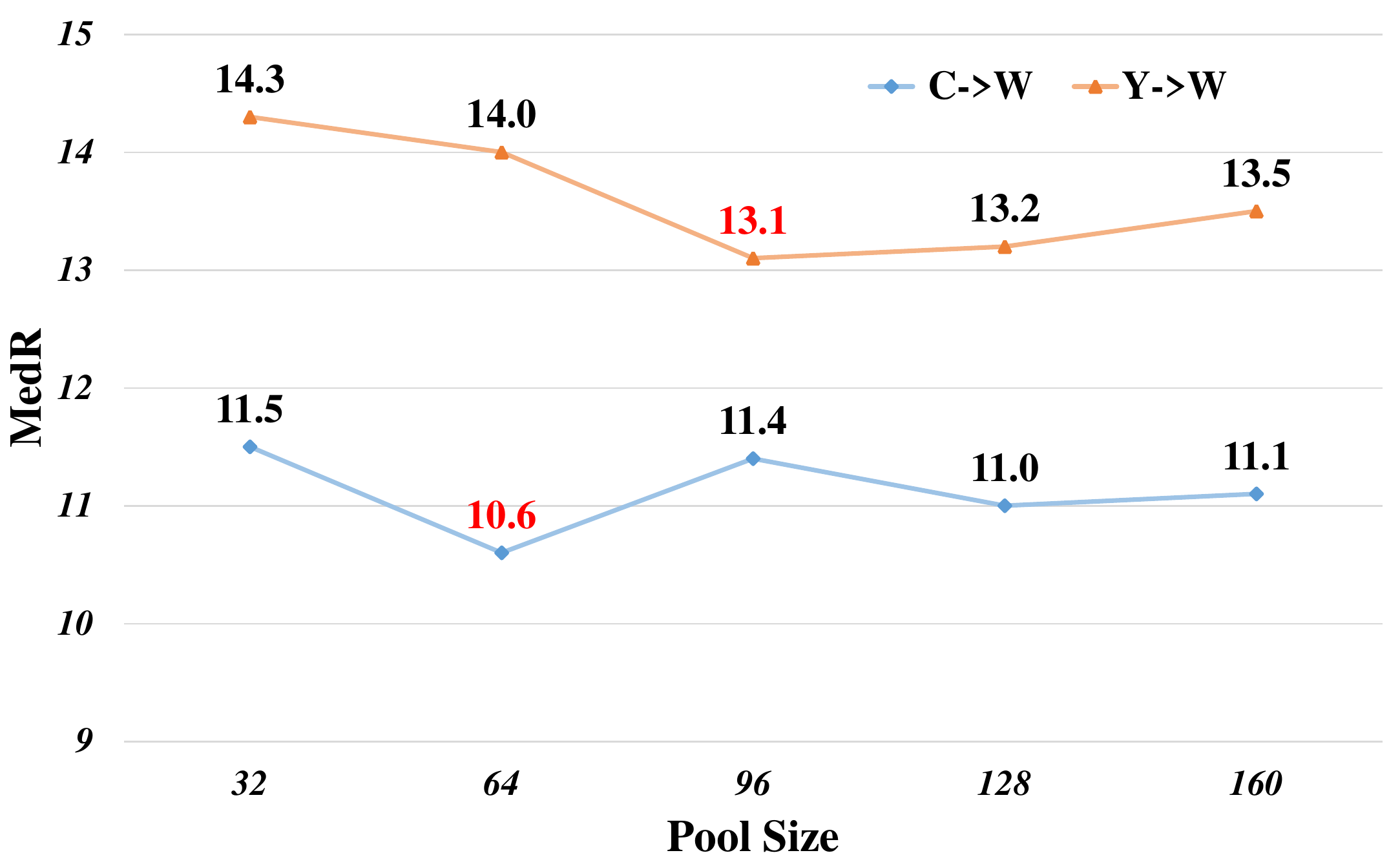}
   \end{center}
      \caption{The MedR performance by varying the size of source subset pool for C$\rightarrow$W and Y$\rightarrow$W. The optimal MedR is highlighted in red.}
   \label{fig:poolsize}
\end{figure}

\textbf{Parameter Selection.} To investigate the impact of source pool size in SBS, the size is varied from one to five times of the training batch size, i.e., \{32, 64, 96, 128, 160\}. Note that when pool size equals to
32, no SBS is employed. The MedR performances of C$\rightarrow$W
and Y$\rightarrow$W are shown in Figure~\ref{fig:poolsize}. Obviously, a larger pool size does not necessarily lead to better performance and the optimal pool size for different cuisine transfers
may differ. As it can be seen, the best pool sizes for C$\rightarrow$W and Y$\rightarrow$W are 64 and 96 respectively. Furthermore, the pool size has a dramatic effect on the performance. Take C$\rightarrow$W
for example, if the pool size is set to be 96 rather than 64, the MedR performance will drop from 10.6 to 11.4, which is almost the same with no usage of SBS with MedR 11.5.

In general, the chance for a target sample to select similar source samples is proportional to the pool size. The effect of SBS could be no difference from random sampling if the pool size is small. On the
other hand, the same set of similar samples could be repeatedly selected when the pool size is large. This will consequently hinder learning efficiency due to the recurrent use of the same or
similar source samples for training. We hence speculate that the selection of pool size is a trade-off between filtering irrelevant source samples from training and maintaining sample
diversity. A moderate pool size, besides being capable of selecting suitable samples to bridge the domain shift, can also increase disturbance in training.

\textbf{Ablation Study} experiments three variants of the proposed model by removing SBS, weighted triplet loss and weighted adversarial learning respectively. As shown in Table~\ref{tab:performance}, comparing to CCFT~\cite{zhumm20}, the improvements
for all the transfers are quite tiny if SBS is not in use, i.e., Proposed w/o SBS. Hence, data selection to filter out distinctive source samples in advance is essential for weighted cross-modal adversarial learning.
When weights are omitted for either weighted triplet loss or weighted adversarial loss, the performance is also degraded in all the transfers. For instance, the MedR ranks drop
0.6 and 0.7 for Proposed w/o w\_adv and w/o w\_triplet in C$\rightarrow$W respectively, while only 0.1 for both variants in C$\rightarrow$Y 1K testing. Between them, weighted triplet loss obtains slightly
more gains than weighted adversarial loss when only one of them is adopted.

\section{Conclusion}
We have presented a new method based on data selection and re-weighting schemes for cross-domain food image-to-recipe retrieval, which outperforms the existing approaches in all the transfers on the datasets of three different cuisines.
The empirical studies show that the pre-trained source model is reasonable as an feature extractor to compute the weights for source samples. Through the experiments, the effectiveness of the proposed method
is clearly shown with constant improvements. As for source pool size selection, it is of great significance to balance between filtering out distinctive samples and maintaining proper sample diversity.
Through ablation study, we show that data filtering to exclude dissimilar source samples is important for subsequent weighted adversarial cross-modal learning. In addition, both weighted triplet loss and
weighted adversarial loss are capable to boost performance. While encouraging, the proposed method still shows limited ability to further improve the performance for fine-grained recipes in the top ranked retrieved recipes, which will
be our future work.

{\small
\bibliographystyle{ieee_fullname}
\bibliography{WAFT}
}

\end{document}